\documentclass{elsart}
\usepackage{graphicx}
\usepackage{amsmath}
\usepackage[german,english]{babel}
\begin{document}
\sloppy
\begin{frontmatter}
\title{Assessing Interaction Networks with Applications to Catastrophe Dynamics and Disaster Management}
\author[a]{Dirk Helbing}
\author[a]{and Christian K\"uhnert}
\address[a]{Institute for Economics and Traffic, and Faculty of Mathematics and\\ Natural Sciences, 
Dresden University of Technology, D-01062 Dresden, Germany}

\begin{abstract}
In this paper we present a versatile method for the investigation
of interaction networks and show how to use it to assess
effects of indirect interactions and feedback loops. The method
allows to evaluate the impact of optimization measures or
failures on the system. Here, we will apply it to the investigation of
catastrophes, in particular to the temporal development of disasters
(catastrophe dynamics). The mathematical methods are related to the
master equation,  which allows the application of well-known solution methods.
We will also indicate connections of disaster management with
excitable media and supply networks. This facilitates to study the effects
of measures taken by the emergency management or the local
operation units. With a fictious, but more or less realistic
example of a spreading epidemic disease or a wave of influenza, 
we illustrate how this method can, in principle, provide decision support to the emergency
management during such a disaster. Similar considerations may help to
assess measures to fight the SARS epidemics, although immunization is presently not
possible.
\end{abstract}

\end{frontmatter}

\section{Introduction}

Natural disasters \cite{catastrophes} have occured since earliest times, and despite
the development of science and technology, they still cause many
victims each year. One reason for this is the increased world population.
Nowadays, more people can be affected by a disaster than in
former centuries. Apart from this, more people affect their environment,
which challenges its stability and triggers disasters such as famines.
\par
One common feature of many disastrous events is
the so-called domino or avalanche effect. It means, that one critical situation
triggers another one and so one, so that the situation worsens even more.
One famous example is a mountain slide that fell into a lake and caused very
high waves (Vajont, Italy 1963 \cite{vajont}). Other examples are fires and failures of
water and electricity supply caused by earthquakes or, on a larger timescale,
a disease harming the economy and social life of the
affected area, which leaves the people and country even poorer and the medical
system less effective, causing further fatalities (e.g. AIDS in
Central Afrika \cite{aids} or the plague in former ages).
\par
Therefore, a great effort is necessary to prevent the emergence of disasters
(which is not always possible) and to improve the management of catastrophes.
Physics and other natural scientists have traditionally contributed a lot to
understanding the laws behind catastrophes. For example, we mention the
extensive work on forest fires \cite{fire} and earthquakes \cite{earthquake},
which relate to concepts of self-organized criticality
used to describe avalanche effects \cite{SOC}. But there is also work on floodings
\cite{flooding}, landslides \cite{landslide}, and volcanos \cite{volcano}.
A considerably attention has also been devoted to epidemics \cite{epidemic}.
\par
Some of the physical disciplines involved into the study of these subjects are
the theory of catastrophes and bifurcations \cite{catastrophe},
non-equilibrium phase transitions \cite{phase}, self-organized criticality
and scaling laws \cite{SOC}, percolation theory \cite{percolation},
the statistical physics of networks \cite{nets} and extreme events \cite{extreme},
stochastic processes \cite{stoch} and noise-induced transitions \cite{noise}, but mechanics, fluid-dynamics,
and other fields play an important role as well \cite{math}.
\par
In this paper, we like to develop a flexible semi-quantitative method
allowing
\begin{itemize}
\item to assess the suitability of alternative measures of
emergency management, i.e. to give decision support,
\item to estimate the temporary development of catastrophes, and
\item to give hints when to take certain actions in an anticipatory way.
\end{itemize}
For that purpose, it is necessary to take into account all factors which are relevant
during the catastrophe and all direct and indirect interactions
between them. This method is in the tradition of systems theory \cite{SystDyn}.
It extends the concept of the causality diagram in Sec.~\ref{Kausal},
while a dynamical generalization is developed in Sec.~\ref{Dyn}.
\par
In the next section, we start with a static analysis of interaction networks.
Section~\ref{ex} is intended to illustrate its usefulness for disaster
management. There, we will calculate in a semi-fictious example
the effects of measures  taken to combat an epidemic catastrophe.
Section~\ref{Dyn} contains the extension to a dynamic
description, which is connected to the discrete master equation \cite{stoch}.
It allows to determine the probability and order of events as well as their most
likely occurence in time. In Secs.~\ref{exci} and \ref{supp}, we will develop a dynamical
model of disaster management, which specifies some parameters in the master
equation. It relates to models of excitable media \cite{excite} and supply networks \cite{Production,Supplynet}.
Some analytical results for this model are presented in Sec.~\ref{simu}, while Sec.~\ref{summ} summarizes
our results and closes with an outlook.

\section{Assessment of Interaction Networks} \label{Kausal}

In this section, we want to develop a simple method to reflect
the approximate influence of different factors or sectors on each other.
Such factors may, for example, be energy supply, public transport, or medical
support. In principle, it is a long list of variables $i$, which may play a role
for the problem under consideration. If we represent the influence of
factor $j$ on factor $i$ by $A_{ij}$, we may summarize these (directed) influences
by a matrix $\mathbf{A} = (A_{ij})$. However, in practical applications, one
faces the following problems:
\begin{enumerate}
\item The number of possible interactions grows quadratically with the
number of variables or factors $i$. It is, therefore, difficult to measure
or even estimate all the influences $A_{ij}$.
\item While it appears feasible to determine the {\em direct} influence $M_{ij}$ of one variable
$j$ on another one $i$, it is hard or almost impossible to estimate the indirect influences over
various nodes of the graph, which enter into $A_{ij}$ as well.
However, feedback loops may have an important effect and
may neutralize or even overcompensate the direct influences.
\end{enumerate}
Problem (i) can be partially resolved by clustering similar variables and
selecting a representative one for each cluster. The remaining set of variables should contain
the main explanatory variables. Systematic statistical methods for such a procedure are,
in principle, available, but intuition may be a good guide, when the quantitative
data required for the clustering of variables are missing.
\par
Problem (ii) may be addressed by estimating the {\em indirect} influences due to feedback
loops based on the {\em direct} influcences $M_{ij}$, which can be summarized by a matrix
$\mathbf{M} = (M_{ij})$. One may use a formula such as
 \begin{equation}
 \mathbf{A}' = \mathbf{A}'_\tau = \frac{1}{\tau} \sum_{k=1}^\infty (\tau \mathbf{M})^k 
 = \frac{1}{\tau} \sum_{k=1}^\infty \tau^k \mathbf{M}^k
 = \sum_{k=1}^\infty \tau^{k-1} \mathbf{M}^k \, ,
\end{equation}
but as this converges only for small enough values of $\tau$, we will instead use the formula
\begin{equation}
 \mathbf{A} = \mathbf{A}_\tau =\frac{1}{\tau} \sum_{k=1}^\infty \frac{\tau^k \mathbf{M}^k}{k!}
 = \frac{1}{\tau} [ \exp(\tau \mathbf{M}) - \mathbf{1} ] \, ,
\label{formula}
\end{equation}
where $\mathbf{1}$ denotes the unity matrix. The expression $\mathbf{M}^k$ reflects
all influences over $k-1$ nodes and $k$ links, i.e. $k=1$ corresponds to direct influences, $k=2$ to feedback loops
with one intermediate node, $k=3$ to feedback loops with two intermediate nodes, etc. The prefactor $\tau^k$
is not only required for convergence, but with $\tau < 1$, it also allows to reflect that indirect interactions
often become weaker, the more edges (nodes) are in between.
\par
A further simplification can be reached by restricting to a few discrete values
to characterize the influences. We may, for example, restrict ourselves to
\begin{equation}
  M_{ij} \in \{-3,-2,-1,0,1,2,3\} \, ,
\end{equation}
where $M_{ij} = \pm 3$ means an extreme positive or negative influence,
$M_{ij} = \pm 2$ represents a strong influence, $M_{ij} = \pm 1$ a weak
influence, and $M_{ij} = 0$ a negligible influence. Of course, a finer differentiation
is possible, whenever necessary. (For an investigation of stylized relationships,
it can also make sense to choose $M_{ij} \in \{-1,0,1\}$, where 
$M_{ij} = \pm 1$ represents a strong positive or negative influence, then.) 
\par
The matrix $\mathbf{A} = (A_{ij})$ will be called the assessment matrix and summarizes all direct
influences ($\mathbf{M}$) and feedback effects ($\mathbf{A} - \mathbf{M}$)
among the investigated factors. It allows conclusions about 
\begin{itemize}
\item the resulting strength of desireable and undesireable interactions, when feedback
effects are included,
\item the effect of failures of a specific sector (node),
\item the suitability of possible measures to reach specific goals or improvements,
\item the side effects of these measures on other factors.
\end{itemize}
This will be illustrated in more detail by the example in Sec.~\ref{ex}.
\par
One open problem is the choice of the parameter $\tau$. It controls how strong the
indirect effects contribute in comparison with the direct effects. A small value of
$\tau$ corresponds to neglecting indirect effects, i.e.
\begin{equation}
 \lim_{\tau \rightarrow 0} \mathbf{A}_\tau = \mathbf{M} \, ,
\end{equation}
while increasing values of $\tau$
reflect a growing influence of indirect effects. This is often the case for catastrophes,
as these are frequently related to bifurcations or phase transitions, to avalanches or
percolation effects \cite{earthquake,fire}.
By variation of $\tau$, one can study different scenarios.
\par
Note that $\tau$ may be interpreted as time coordinate: Defining
\begin{equation}
 \vec{X}(\tau) = \exp( \tau \mathbf{M} ) \vec{X}
\end{equation}
for an arbitrary vector $\vec{X}$, we find $\vec{X}(0) = \vec{X}$,
\begin{equation}
 \frac{\vec{X}(\tau)-\vec{X}(0)}{\tau}  =\frac{1}{\tau} [  \exp(\tau \mathbf{M}) - \mathbf{1} ] \vec{X}(0) \nonumber \\
 = \mathbf{A}_\tau \vec{X}(0)
\end{equation}
and
\begin{equation}
 \frac{d\vec{X}}{d\tau} = \lim_{\tau \rightarrow 0} \frac{\vec{X}(\tau)-\vec{X}(0)}{\tau} \nonumber \\
 =  \mathbf{M} \vec{X}(0) \, .
\label{rough}
\end{equation}
From this point of view,
\begin{equation}
   \vec{X}(\tau) = (\tau \mathbf{A}_\tau + \mathbf{1}) \vec{X}(0)
\label{because}
\end{equation}
describes the state of the system at time $\tau$, and
$M_{ij}$ the changing rates. $\vec{X} = \vec{0}$ is a stationary solution and corresponds to the
normal (everyday) state. An initial state $\vec{X}(0) \ne \vec{0}$ may be interpreted as perturbation of the system
by some (catastrophic) event. We should, however, note that the linear system of equations (\ref{rough})
is certainly a rough description of the system dynamics. It is expected to hold only for
small perturbations of the system state and does not consider damping effects due to disaster management.
Such aspects will be considered later on (see Secs.~\ref{Dyn} and \ref{supp}), 
after discussion of an example illustrating how to apply
interaction matrices to cope with catastrophes.

\section{Optimization of Interaction Networks: A Simple Example \label{ex}}

One advantage of our semi-quantitative approach to catastrophes is that it
allows to estimate the impact of certain actions on
the whole set of factors. Usually, during a disaster the
responsibilities have only dissatisfactory information and
short time to decide, so in many cases they will take into account
only direct impacts on other factors.
In the worst case, this may lead to the opposite than the desired
result, if the feedback effects exceed the direct influence.
Therefore, it would be better to know the
implications on the whole system. As we have argued before,
all direct and indirect effects are summarized by the matrix $\mathbf{A}$,
which is determined from the matrix $\mathbf{M}$ of direct interactions.
Different measures taken by the responsibilities  are reflected by different matrices $\mathbf{M}$.
\par
\begin{figure}[h!]
\begin{center}
\includegraphics[scale=.4]{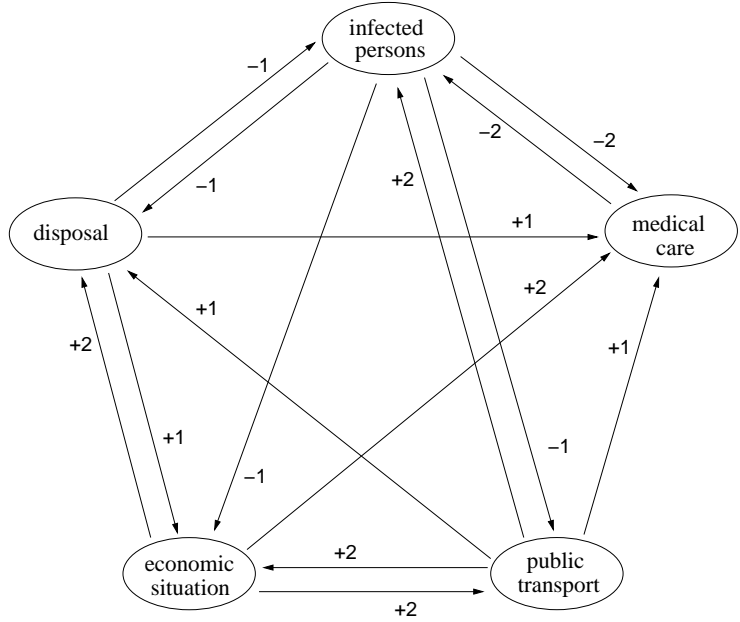}
\end{center}
\caption{Interaction network for the example of a spreading disease discussed in the text.}
\label{disease}
\end{figure}

As an example we consider the spreading of a disease. For illustrative reasons, we
will restrict to the discussion of five factors only:
\begin{enumerate}
\item the number of {\em infected persons},
\item the quality of {\em medical care},
\item the {\em public transport},
\item the {\em economic situation} and
\item the {\em disposal} of waste.
\end{enumerate}
These factors are not independent from each other, as illustrated by
Fig.~\ref{disease}.
\par
The corresponding matrix of the assumed direct influences among the different
factors is
\begin{equation}
\mathbf{M} =
\begin{pmatrix}
0 & -2 & +2 & 0 & -1 \\
-2 & 0 & +1 & +2 & +1  \\
 -1 & 0 & 0 & +2 & 0 \\
 -1 & 0 & +2 & 0 & +1 \\
-1 & 0 & +1 & +2 & 0
\end{pmatrix}
\label{matrixm}
\end{equation}
The right choice of the sign of the direct influence $M_{ij}$ of factor $j$ on factor
$i$ is plausible: We assume a positive sign if the factor $i$ increases with an increase
of factor $j$, while we assume a negative sign when factor $i$ decreases with the growth of
factor $j$. However, the determination of the absolute value of $M_{ij}$ requires
empirical data, expert knowledge, or experience. We have argued as follows:
\begin{itemize}
\item A growing number of infected persons affects all other factors in a negative way (see first column), as
these will not continue to work. That is, there will be problems to maintain a good economic situation,
public transport, or the disposal of waste. Health care is affected twice, as not only the medical
personnel may be infected, but also a higher number of patients needs to be treated, and capacities
are limited. Therefore, we have chosen the value $-2$ in this case, but $-1$ for the other factors.
\item A well operating health system (second column) can reduce the number of infected persons efficiently,
so that we have chosen a value of $-2$. The influence of the health system on the economic situation
and other factors was assumed to be of indirect nature, by reducing the number of ill persons.
\item Public transport (third column) contributes to a fast spreading of the infection assumed here. Therefore,
we have selected a value of $2$. Transport is also an important factor for economic prosperity
(therefore the value of $2$),  and it is required to get medical personnel and workers in the disposal
sector to work (which is reflected by a value of $1$).
\item The economic situation (fourth column) has a significant effect on the quality of the
health system, public transport, and disposal, so that we have chosen a value of $2$ in each case.
\item Waste may contribute to the spreading of the disease, if it is not properly removed. Therefore,
a good disposal system (fifth column) may reduce the number of infections (therefore the value of $-1$). It is also
required for a functioning health system and economic production. That is, why we have assumed a
value of 1.
\end{itemize}
Depending on the respective situation, the concrete values of the direct influences
$M_{ij}$ may be somewhat different. For their specification, it can also be helpful to
check the resulting values of $A_{ij}$ of the overall direct plus indirect influence for their plausibility,
and to  compare the size of second-order or third-order interactions. For example, we see that
the third-order feedback loop ``number of infected persons$\rightarrow$economic situation$\rightarrow$quality
of the health system$\rightarrow$number of infected persons'' is proportional to $(-1)\cdot (+2) \cdot (-2)
= 4$. The same indirect influence is found for the feedback loop
``number of infected persons$\rightarrow$economic situation$\rightarrow$public
transport$\rightarrow$number of infected persons''. Moreover, according to our assumptions,
the second-order autocatalytic increase of the number of infected persons
via its impact on the health system is four times as large as the one via its impact on the waste disposal.
One surprising observation is that the number of infected persons is reduced via its impact on public
transport. In fact, once less busses are operated (because the bus drivers are ill), the spreading rate of
the disease is reduced. This may inspire responsibilities to reduce public transport or even stop it.
Later on, we will discuss the effect of this possible measure.
\par
Before, let us have a look at the resulting overall interaction matrix
\begin{equation}
\mathbf{A} = (A_{ij}) =
\begin{pmatrix}
0.9 & -2.2 & 1.3 & -0.8 & -1.6 \\
-3.4 & 1.1 & 1.5 & 3.5 & 2.3  \\
-1.7 & 0.6 & 0.5 & 2.5 & 0.8 \\
-2.0 & 0.6 & 2.1 & 1.5 & 1.6 \\
-2.0 & 0.6 & 1.5 & 2.9 & 0.9
\end{pmatrix}
\label{ausgang}
\end{equation}
For its calculation, we have chosen the value $\tau=0.4$, which will also be used later on to assess
alternative actions to fight the spreading of the disease. In order to discuss a certain scenario,
we will assume that $X_j$ reflects the perturbation of factor $j$. Because of Eq.~(\ref{because}), the quantities
\begin{equation}
 Y_i = \sum_j (\tau A_{ij} + \delta_{ij}) X_j
\label{quant}
\end{equation}
will be used to characterize the potential
response of the system in the specific scenario described by the perturbations $X_j$ (and without the
damping effects by disaster management discussed in later sections of this contribution).
Here, $\delta_{ij}$ denotes the Kronecker function, which is 1 for $i=j$ and 0 otherwise.
We will assume $X_1 = 1.0$, as the number of infected persons is higher than normal, and
$X_2 = X_3 = X_4 = X_5 = -0.1$, as the other factors are reduced by the spreading of the disease:
\begin{equation}
 (X_1,X_2,X_3,X_4,X_5) = (1.0,-0.1,-0.1,-0.1,-0.1) \, .
\end{equation}
Moreover, if we attribute a weight of
$Z_1=0.5$ to the number of infected persons, a weight $Z_4 = 0.3$ to the economic
situation, and weights of $Z_2=Z_3 =0.1$ to the quality of the medical care and public transport,
while we do not care about waste in our evaluation (i.e. $Z_5 = 0$), the resulting value of
\begin{equation}
 F = F_\tau = \left( \sum_i Z_i Y_i^2 \right)^{1/2}
\end{equation}
will be used to assess the overall situation of the system. In the stationary (normal) system state,
$F$ would be zero. Therefore, we want to find a strategy which brings $F$ close to zero. 
For our basic scenario, we find
\begin{equation}
 (Y_1, Y_2, Y_3,Y_4,Y_5) = (1.5, -1.8, -1.0, -1.1, -1.1) \quad \mbox{and} \quad
 F = 1.4 \, .
\end{equation}
These reference values will be compared with the values for alternative
scenarios which correspond to different actions taken to fight the catastrophe.
\par
For example, let us assume to have limited stocks of vaccine for immunization. Should we use these
to immunize 1) the transport workers, 2) the medical staff, or 3) the disposal workers? In the first
case, we have the modified matrix
\begin{equation}
\mathbf{M} =
\begin{pmatrix}
0 & -2 & +2 & 0 & -1 \\
-2 & 0 & +1 & +2 & +1  \\
\underline{0} & 0 & 0 & +2 & 0 \\
 -1 & 0 & +2 & 0 & +1 \\
-1 & 0 & +1 & +2 & 0
\end{pmatrix}  \, ,
\end{equation}
which implies
\begin{equation}
\mathbf{A} =
\begin{pmatrix}
1.2 & -2.3 & 1.4  & -0.8 & -1.7 \\
-3.2 & 1.1 & 1.5  &  3.5 &  2.2  \\
-0.5 & 0.1 & 0.8  &  2.4 &  0.5 \\
 -1.6 & 0.5 & 2.2 &  1.5 &  1.5 \\
-1.7 & 0.6 & 1.6  &  2.9 &  0.9
\end{pmatrix} \, ,
\end{equation}
\begin{equation}
 (Y_1, Y_2, Y_3,Y_4,Y_5) = (1.6, -1.7, -0.5, -1.0, -1.0)\, ,  \quad \mbox{and} \quad
 F = 1.4 \, .
\end{equation}
In the second case, when we immunize the medical staff, we find
\begin{equation}
\mathbf{M} =
\begin{pmatrix}
0 & -2 & +2 & 0 & -1 \\
\underline{-1} & 0 & +1 & +2 & +1  \\
 -1 & 0 & 0 & +2 & 0 \\
 -1 & 0 & +2 & 0 & +1 \\
-1 & 0 & +1 & +2 & 0
\end{pmatrix}  \, ,
\end{equation}
which implies
\begin{equation}
\mathbf{A} =
\begin{pmatrix}
0.5   & -2.1 & 1.3 & -0.7 & -1.5 \\
-2.3  & 0.7 & 1.8 & 3.4 & 2.0  \\
-1.7  & 0.6 & 0.5 & 2.5 & 0.8 \\
-1.9  & 0.6 & 2.1 & 1.5 & 1.6 \\
-1.9  & 0.6 & 1.6 & 2.9 & 0.9
\end{pmatrix} \, ,
\end{equation}
\begin{equation}
(Y_1, Y_2, Y_3,Y_4,Y_5) = (1.3, -1.3, -0.9, -1.1, -1.1) \, , \quad \mbox{and} \quad
 F = 1.2 \, .
\end{equation}
In the third case, when the disposal workers are immunized, we expect
\begin{equation}
\mathbf{M} =
\begin{pmatrix}
0 & -2 & +2 & 0 & -1 \\
-2 & 0 & +1 & +2 & +1  \\
 -1 & 0 & 0 & +2 & 0 \\
 -1 & 0 & +2 & 0 & +1 \\
\underline{0} & 0 & +1 & +2 & 0
\end{pmatrix}  \, ,
\end{equation}
which implies
\begin{equation}
\mathbf{A} =
\begin{pmatrix}
0.6 & -2.1 & 1.3 & -0.7 & -1.6 \\
-3.1 & 1.1 & 1.6 & 3.5 & 2.2  \\
-1.6 & 0.6 & 0.5 & 2.5 & 0.8 \\
-1.7 & 0.6 & 2.1 & 1.5 & 1.5 \\
-0.8 & 0.2 & 1.9 & 2.8 & 0.6
\end{pmatrix} \, ,
\end{equation}
\begin{equation}
 (Y_1, Y_2, Y_3,Y_4,Y_5) = (1.4, -1.7, -0.9, -1.0, -0.7) \, , \quad \mbox{and} \quad
 F = 1.3 \, .
\end{equation}
While the immunization of the public transport staff has almost no effect on the
overall situation in the system, the last two measures can improve it. We see that
it is more effective to immunize the medical staff than the disposal workers,
and the best would be to immunize both groups. This corresponds to
\begin{equation}
\mathbf{M} =
\begin{pmatrix}
0 & -2 & +2 & 0 & -1 \\
\underline{-1} & 0 & +1 & +2 & +1  \\
 -1 & 0 & 0 & +2 & 0 \\
 -1 & 0 & +2 & 0 & +1 \\
\underline{0} & 0 & +1 & +2 & 0
\end{pmatrix}  \, ,
\label{immunization}
\end{equation}
and we obtain
\begin{equation}
\mathbf{A} =
\begin{pmatrix}
0.2 & -2.0 & 1.2 & -0.7 & -1.5 \\
-1.9 & 0.6 & 1.9 & 3.4 & 1.9  \\
-1.6 & 0.5 & 0.5 & 2.5 & 0.8 \\
 -1.7 & 0.6 & 2.1 & 1.5 & 1.5 \\
-0.8 & 0.2 & 1.9 & 2.8 & 0.6
\end{pmatrix} \, ,
\end{equation}
\begin{equation}
 (Y_1, Y_2, Y_3,Y_4,Y_5) = (1.2, -1.2, -0.9, -1.0, -0.6) \, , \quad \mbox{and} \quad
 F = 1.1 \, .
\end{equation}
\par
Other measures do not change the interactions in the system, but correspond to a change 
of the effective impact $\vec{X}$
of the catastrophe. For example, we may consider to reduce public transport.
With (\ref{ausgang}) and 
\begin{equation}
 (X_1,X_2,X_3,X_4,X_5) = (1.0,-0.1,\underline{-1.0},-0.1,-0.1) \, ,
\end{equation}
we find
\begin{equation}
 (Y_1, Y_2, Y_3,Y_4,Y_5) = (1.0, -2.4, -2.0, -1.9, -1.7) \quad \mbox{and} \quad
 F = 1.6 \, .
\end{equation}
We see that the number of infections could, in fact, be reduced. However, the overall situation
of the system has deteriorated, as the economic situation and all the other sectors were negatively
affected, because many people could not reach their workplace. Therefore, let us consider
the option to increase the number of disposal workers. With (\ref{ausgang}) and
\begin{equation}
(X_1,X_2,X_3,X_4,X_5) = (1.0,-0.1,-0.1,-0.1,\underline{0.5}) \, ,
\label{disposal}
\end{equation}
we find
\begin{equation}
 (Y_1, Y_2, Y_3,Y_4,Y_5) = (1.1, -1.3, -0.8, -0.8, -0.3) \quad \mbox{and} \quad
 F = 1.0 \, .
\end{equation}
In conclusion, increasing the hygienic standards can be surprisingly efficient. \\
\par
Finally, let us assume an improved waste disposal together
with the immunization of both, the medical staff and the disposal
workers. In that case the interactions of the relevant factors 
are characterized by matrix (\ref{immunization}), whereas the
starting vector is again (\ref{disposal}). The resulting response
is
\begin{equation}
(Y_1, Y_2, Y_3,Y_4,Y_5) = (0.8, -0.7, -0.7, -0.6, 0.1) \quad \mbox{and} \quad
F = 0.75 \, .
\end{equation}
Only this combination of measures manages to actually reduce the infections compared to
the initial state, i.e. $Y_1 < X_1$. However, we can also see that a negative impact
on the economic situation and other factors cannot be avoided. In any case, we can
assess which measures are reasonable to take, which impact they will have on the system,
and which measures need to be combined in order to control the spreading of the disease
(or other problems in different scenarios).
\par
The simple example in this section was chosen to illustrate the procedure how to assess
interaction networks and potential optimization measures. In an on-going project,
we do now investigate the interaction network among a large number of factors for the floodings in Germany
during August 2002 and other catastrophes. This involves considerably more detailed and
much larger matrices $\mathbf{M}$, where nobody would be able to assess the
feedback loops without a method such as the one proposed above. It also involves other
aspects such as human forces fighting the catastrophe and the availability of technical
or other equipment etc, which will be modeled in the following sections.

\section{Impact of the Interaction Network on Catastrophe Dynamics}\label{Dyn}

Before, we have used the interaction network predominantly for the static assessment of the
influence of different factors on each other.
We will now try to extend this method step by step in a way that allows
a semi-quantitative analysis of the time-dependence of catastrophes for the purpose
of anticipation, which helps to prepare for the next step in catastrophe management or prevention.
We are particularly interested in the domino or avalanche
effects of particular events such as the failure of a certain factor or sector
in the interaction network. We will assume that this failure spreads along and
in the order of the direct connections in the interaction network (causality graph).
In terms of example in Sec.~\ref{ex},
a failure of medical care would first affect the number of infected persons, 
and in a second step the economic situation, public transport, and the disposal of waste.
\par
For a description of the catastrophe dynamics, let us assume that $P_i(\tau)$ denotes the impact 
on factor $i$ at time $\tau$ and $W_{ji}$ the rate at which this impact spreads to factor $j$,
while $D_i$ is a damping rate describing
the mitigation of the catastrophic impact on factor $i$ by disaster management.
In this case, it is reasonable to assume the dynamics
\begin{equation}
 \frac{d\vec{P}}{d\tau} = (\mathbf{W} - \mathbf{D}) \vec{P}(\tau) = \mathbf{L} \vec{P}(\tau)
\label{DGL}
\end{equation}
with $\mathbf{D}=(\delta_{ij} D_i)$, $\mathbf{L} = (L_{ij}) = (W_{ij} - \delta_{ij} D_i)$, and 
$\vec{P}(\tau) = (P_i(\tau))$. 
The symbol $\delta_{ij}$ represents the Kronecker function, i.e. it is 1 for $i=j$ and otherwise
0. When no better information is available, we may assume that the spreading rate $W_{ij}$ is proportional
to the strength $|M_{ij}|$ of the direct influence of factor $j$ on factor $i$. With a constant
proportionality factor $c$ this means
\begin{equation}
 W_{ij} \approx c |M_{ij}| \, .
\label{approx}
\end{equation} 
The formal solution of equation (\ref{DGL}) for a time-independent
matrix $\mathbf{L}$ is given by
\begin{equation}
 \vec{P}(\tau) = \exp(\mathbf{L}\tau) \vec{P}(0)
= \sum_{k=0}^\infty \frac{\tau^k}{k!} \mathbf{L}^k \vec{P}(0)
= \mathbf{B}(\tau) \vec{P}(0) \, .
\end{equation}
That is, $\mathbf{B}(\tau)$ describes the spreading of an event in the causality network (interaction network)
in the course of time $\tau$, while $\vec{P}(0)$ reflects the initial impact of a catastrophic
event. A random series of catastrophic events can be described by adding a random
variable $\vec{\xi}(t)$ to the right-hand side of Eq.~(\ref{DGL}).  
\par
When we assume
\begin{equation}
 D_i = \sum_j W_{ji} \, ,
\label{case}
\end{equation}
equation (\ref{DGL}) is related to the Liouville representation of the discrete master equation. 
In this case, we can apply all the solution methods developed for it. This includes the 
so-called path integral solution \cite{Path}, which allows one to calculate the occurence probability of
specific spreading paths. This has some interesting implications. For example, the danger that
the impact on sector $i_0$ affects the sectors $i_1, i_2, \dots, i_n$ in the indicated order
is quantified by
\begin{equation}
 P(i_0\rightarrow i_1 \rightarrow \dots \rightarrow i_n)
 = \frac{|P_{i_0}(0)|}{D_{i_n}} \prod_{l=0}^{n-1} \frac{W_{i_{l+1},i_{l}}}{D_{i_l}} 
 \approx  c^n \frac{|P_{i_0}(0)|}{D_{i_n}} \prod_{l=0}^{n-1} \frac{|M_{i_{l+1},i_{l}}|}{D_{i_l}} \, .
\label{eins}
\end{equation}
Moreover, the average time at which this series of events has occured can be calculated as
\begin{equation}
  T(i_0\rightarrow i_1 \rightarrow \dots \rightarrow i_n)
 = \sum_{l=0}^n \frac{1}{D_{i_l}} \, ,
\label{zwei}
\end{equation}
and the variance of this time is determined by
\begin{equation}
   \Theta(i_0\rightarrow i_1 \rightarrow \dots \rightarrow i_n)
 = \sum_{l=0}^n \frac{1}{(D_{i_l})^2} \, .
\label{drei}
\end{equation}
That is, Eq. (\ref{DGL}) does not only allow to assess the likelyhood of certain series of
events rather accurately, but also their approximate appearance times. In other words,
we have a detailed picture of the potential catastrophic scenarios and of their
time evolution, which allows for a specific preparation and disaster management.
\par
In the following, we do not want to restrict to the case (\ref{case}). If
\begin{equation}
 D_i < \sum_j W_{ji}
\end{equation}
for all $i$, the damping is weak and the solutions $P_i(\tau)$ are expected to grow
more or less exponentially in the course of time, which describes a scenario where control
is lost and the catastrophe spreads all over the system. In many cases, we will have
\begin{equation}
 D_i > \sum_j W_{ji}
\end{equation}
for all $i$, i.e. the impact of the catastrophe on the system decays in the course of time,
and $\lim_{\tau \rightarrow 0} P_i(\tau) \rightarrow 0$. This determines, how strong the
damping effects, i.e. the means counteracting the catastrophe have to be chosen. In terms of
Sec.~\ref{supp}, this concerns the specification of the parameters $V_{ik}$.
\par
Finally, it may also happen that
$D_i > \sum_j W_{ji}$ for some factors $i$, but $D_i < \sum_j W_{ji}$ for others. In
such situations, everything depends on the initial impact $\vec{P}(0)$ and on the 
matrix $\mathbf{B}(\tau)$. 
However, in all these cases, Eqs.~(\ref{eins}) to (\ref{drei}) remain valid.

\section{An Excitable Media Model of Disaster Management} \label{exci}

The damping effects $D_i$ are, to a large extent, related to the forces 
counter-acting the catastrophe. Therefore, we will now develop a dynamical model for these,
while our previous considerations assumed some more or less constant value of $D_i$.
Let us denote by $N_k$ the quantity of human forces (e.g. police, fire fighers, or military)
ready for action, or the quantity of materials (e.g. technical or medical equipment)
ready for use to fight the catastrophe. The index $k$ distinguishes different kinds
of forces or required materials. We will assume the following equation:
\begin{equation}
 \frac{dN_k}{d\tau} = \frac{R_k(\tau)}{T_k^R} \pm \lambda_k N_k^\pm \mbox{e}^{-\lambda_k \tau}
- \sum_i | P_i | V_{ik} N_k(\tau) A_k{}^l(\tau)\, .
\end{equation}
The first term on the right-hand side describes the quantity $R_k$ of (human)
forces and materials, which were exhausted or
not usable, but become available again after an average time period of $T_k^R$. The second term
delineates reserve forces of quantity $N_k^\pm$, which are activated from the ``standby mode'' at a rate $\lambda_k$
after the occurence of a catastrophe (plus sign), while they are removed after
the recovery from the disaster (minus sign). In most cases, $N_k^- \le N_k^+$, due to possible fatalities.
The third term describes the activation of the forces $k$ to fight the problems with factor or sector
$i$. For simplicity, it is here assumed to be proportional to the strenth $|P_i|$ of the catastrophic impact 
on factor $i$, with proportionality factors $V_{ik}$ which reflect the priorities in disaster management
and the speed with which the forces or materials $k$ become available for $i$. 
The exponent $l$ allows to distinguish different cases: When the impact of the catastrophe on factor
$i$ is known, we may assume $l=0$. However, when the active forces are assumed to order more forces,
an exponent $l>0$ can make sense as well. 
\par
The quantities $A_k$ of forces in action or materials in use change in time according to the equation
\begin{equation}
 \frac{dA_k}{d\tau} = \sum_i |P_i|  V_{ik} N_k(\tau) A_k{}^l(\tau) - \frac{A_k(\tau)}{T_k^A} 
-  \sum_i |P_i| \nu_{ik} A_k(\tau) \, .
\end{equation}
The first term on the right-hand side is due to the available forces $k$, which are activated for
fighting the catastrophe, while the second term describes a reduction of the active forces,
as they become exhausted or damaged and require rest or repair after an average time period
of $T_k^A$. The third term describes unrecoverable losses such as fatalities or unrecoverable
damage of materials, which are assumed to occur with a rate $|P_i| \nu_{ik}$ proportional to the
catastrophic impact $|P_i|$.
\par
The quantity of exhausted or damaged forces is described by the following differential equation:
\begin{equation}
 \frac{dR_k}{d\tau} = \frac{A_k(\tau)}{T_k^A} - \frac{R_k(\tau)}{T_k^R} \, .
\end{equation}
Herein, $1/T_k^A$ is the rate at which the forces $k$ become exhausted or damaged, while
$1/T_k^R$ is the recovery or repair rate.
\par
The above model already contain many effects which are typically relevant in practical 
situations. We should note that it is related to models of excitable media developed to
describe chemical waves, the propagation of electrical pulses in heart tissue, 
LaOla waves in human crowds in stadia \cite{excite}, or the spreading of forest fires \cite{fire}.
These models typically contain three different states: an excitable one, an active one, and a refractory one.
In our model of disaster management, refractory states are described by the variables $R_k$,
active states by the variables $A_k$, and excitable states by the variables $N_k$. Due to this
analogy, we expect to find certain pattern formation phenomena for our model of disaster
management. In a forth-coming paper, this aspect shall be investigated in more detail.

\section{Some Analytical Results} \label{simu}

In the following, we will try to get an idea of the possible behavior of the excitable media
model of disaster management suggested in Sec.~\ref{exci}. The stationary state of this model is,
for $\nu_{ik} = 0$, given by 
\begin{equation}
 \frac{R_k}{T_k^R}  = \frac{A_k}{T_k^A} = \sum_i |P_i| V_{ik} N_k A_k{}^l = \mbox{const} \, .
\end{equation}
In order to investigate the sensitivity with
respect to small perturbations, we will carry out a linear stability analysis of the simplified model
with one sector $i$, $|P_i| \approx \mbox{const.}$, $\nu_{ik} = 0$, and one kind $k$ of forces. 
Dropping the subscripts and defining $P = |P_i|$, the resulting coupled set of differential equations is:
\begin{eqnarray}
 \qquad \qquad \qquad \qquad \frac{dN}{d\tau} &=& \frac{R}{T^R} - PVNA^l \, , \\
 \frac{dA}{d\tau} &=& PVNA^l - \frac{A}{T^A} \, , \\
 \frac{dR}{d\tau} &=& \frac{A}{T^A} - \frac{R}{T^R} \, .
\end{eqnarray}
Its stationary solution is given by $N(\tau) = N_0$, $R(\tau) = R_0$, and $A(\tau) = A_0$ with 
\begin{equation}
\frac{R_0}{T^R} = \frac{A_0}{T^A} = PVN_0A_0{}^l  \, .
\label{take}
\end{equation}
It is stable with respect to disturbances, if all eigenvalues $\lambda$ are non-positive. These eigenvalues
can be calculated in the usual way. Assuming
\begin{equation}
 N(t) = N_0 + \delta N \; \mbox{e}^{\lambda \tau}\, , \quad 
 A(t) = A_0 + \delta A \; \mbox{e}^{\lambda \tau}\, , \quad \mbox{and} \quad
 R(t) = R_0 + \delta R \; \mbox{e}^{\lambda \tau}\, , 
\end{equation}
we find the following eigenvalue problem for
the amplitudes $\delta N$, $\delta A$, and $\delta R$ of the deviations
from the stationary values $N_0$, $A_0$, and $R_0$:
\begin{equation}
 \lambda \left( \begin{array}{c}
\delta N \\ \delta A \\ \delta R 
\end{array} \right)
= \left( \begin{array}{ccc}
-PVA_0{}^l & -PVN_0lA_0{}^{l-1} & 1/T^R \\
PVA_0{}^l & PVN_0 lA_0{}^{l-1} - 1/T^A & 0 \\
0 & 1/T^A & - 1/T^R 
\end{array} \right)
\left( \begin{array}{c}
\delta N \\ \delta A \\ \delta R \\
\end{array} \right) \, .
\end{equation}
The eigenvalues $\lambda$ are the solutions of the characteristic equation
\begin{eqnarray}
\qquad \quad & & (-PVA_0{}^l -\lambda)\left(PVN_0lA_0{}^{l-1} - \frac{1}{T^A} - \lambda \right)
\left(-\frac{1}{T^R} - \lambda \right) \nonumber \\
 & & \qquad \qquad + \frac{PVA_0{}^l}{T^R T^A}
+P^2 V^2 N_0 l A_0{}^{2l-1} \left( - \frac{1}{T^R} - \lambda \right) = 0 \, .
\end{eqnarray}
The three solutions are
\begin{eqnarray}
 \qquad \lambda_{1/2} &=& \frac{P V A_0{}^{l-1} (l N_0 - A_0) }{2}  - \frac{1}{2T^A} -
\frac{1}{2T^R} \nonumber \\[2mm]
 &\pm & \sqrt{  \frac{1}{4} \left[ P V A_0{}^{l-1} (l N_0 - A_0) + \frac{1}{T^R} -
\frac{1}{T^A} \right]^2 - \frac{P V A_0{}^l}{T^A}   }
\end{eqnarray}
and $\lambda_3 = 0$. Taking into account Eq.~(\ref{take}), which implies
$PVA_0{}^{l-1}= 1/(N_0 T^A)$, this becomes
\begin{equation}
  \lambda_{1/2} = \frac{1}{2T^A} \left(l-1-\frac{A_0}{N_0}\right) - \frac{1}{2T^R}
 \pm \sqrt{ \left[ \frac{1}{2T^A} \left(l-1-\frac{A_0}{N_0}\right) + \frac{1}{2T^R} \right]^2
 - \frac{A_0}{N_0(T^A)^2} } \, .
\end{equation}
A detailed analysis of this expression shows the following: The system behaves unstable with
respect to perturbations, when the real part of one of the above solutions becomes positive,
which is the case for
\begin{equation}
  l - 1 - \frac{A_0}{N_0} > \min \left( \frac{A_0}{N_0} \frac{
T^R}{T^A} \; , \; \frac{T^A}{T^R} \right) \, .  
\label{instabi}
\end{equation}
Otherwise (apart from the case of marginal stability resulting for the equality sign), 
perturbations are damped, but one can distinguish two subcases: For
\begin{equation}
 -\frac{T^A}{T^R} - 2 \sqrt{\frac{A_0}{N_0}} < l-1 - \frac{A_0}{N_0} < 
\min \left( \frac{A_0}{N_0} \frac{
T^R}{T^A} \; , \; \frac{T^A}{T^R} \; , \; -\frac{T^A}{T^R} + 2 \sqrt{\frac{A_0}{N_0}} \right) \, ,
\end{equation}
the resulting solution is complex, corresponding to damped oscillations, while the 
system behaves overdamped in the remaining case, where perturbations fade away without any oscillations.
For disaster management, the linearly unstable case and the case of damped oscillations
are both unfavourable. Therefore, $l$ should be small enough. Otherwise, 
if active forces recruit other forces, the resulting ``autocatalytic effect'' may cause instabilities or overreactions in
the supply with forces and materials. This effect is most likely for disasters which nobody was prepared for, where
the recruiting mechanism plays the most signifant role. It may explain the suboptimal distribution
of forces observed in these situations \cite{Kirchbach}.

\section{Connection with Supply Networks and  Production Systems} \label{supp}

Finally, we have to specify the influence of disaster mangement activities on the damping $D_i$
of the impact $P_i$, which a catastrophic event has on sector $i$. Let us assume that we have $K$ different kinds
of forces, materials or technical equipment. With $k\in K$, we will indicate that the forces or materials
$k$ can substitute each other, i.e. $K$ summarizes equivalent forces or materials. On the other hand,
certain actions require the simultaneous presence of different supplementary kinds of forces and materials.
Let us assume that the quantities simultaneously required to reduce the problems with factor $i$ are
represented by the coefficients $c_{iK}$. The units of $c_{iK}$
shall be chosen in a way that the following equation holds:
\begin{equation}
 D_i(\tau) = (1- L_i) \min_K \left\{ \frac{\sum_{k\in K} |P_i| V_{ik}N_k(\tau) A_k{}^l(\tau)}{c_{iK}} \right\} \, ,
\label{disaster}
\end{equation}
where $V_{ik}N_k(\tau) A_k{}^l(\tau)$ is the rate of activating forces $k$ to mitigate the situation of factor $i$.
This equation reflects that, if only one of the required forces or materials is missing, no successful
action can be taken. Moreover, there may be a loss $L_i$ of efficiency, e.g. due to queueing or limited capacities
($0\le L_i \le 1$). The formula is analogous to that for production systems, where a product cannot
be finished, as long as some required part or worker is missing, and where finite storage capacities
may cause losses \cite{Production}. Therefore, this formula
delineates the inefficiencies in disaster management, which occur when forces and materials are
distributed in the wrong way. It has sometimes been reported, that too many forces have been
located at some place, and missing at others \cite{Kirchbach}. Here, models designed to optimize supply networks
could help to optimize the efficiency of disaster management \cite{Supplynet}.
\par
Note that, for disaster management, a slight generalization of formula (\ref{disaster}) is in place, as
improvization may cope with a lack of certain materials or forces. It is reasonable to assume
a generalized function $G_q$ with
\begin{equation}
 D_i(\tau) = (1- L_i) G_q\left( \left\{ \frac{\sum_{k\in K}  |P_i| V_{ik}N_k(\tau) A_k{}^l(\tau)}{c_{iK}} \right\} \right)
\label{toge}
\end{equation}
and
\begin{eqnarray}
 \min_K \left\{ \frac{\sum_{k\in K}  |P_i| V_{ik}N_k(\tau) A_k{}^l(\tau)}{c_{iK}} \right\} &\le &
 G_q\left( \left\{ \frac{\sum_{k\in K}  |P_i| V_{ik}N_k(\tau) A_k{}^l(\tau)}{c_{iK}} \right\} \right) \nonumber \\
 &\le & \sum_K \frac{\sum_{k\in K}  |P_i| V_{ik}N_k(\tau) A_k{}^l(\tau)}{c_{iK}} \, .
\end{eqnarray}
Herein, the minimum reflects the worst case, while the sum over $K$ describes the best case (if 
we neglect non-linearities, which may sometimes arise due to synergy effects). The specification
\begin{equation}
  G_q\left( \left\{ \frac{\sum_{k\in K}  |P_i| V_{ik}N_k(\tau) A_k{}^l(\tau)}{c_{iK}} \right\} \right)
 = \left[ \sum_K \left( \frac{\sum_{k\in K}  |P_i| V_{ik}N_k(\tau) A_k{}^l(\tau)}{c_{iK}} \right)^q \right]^{1/q}
\end{equation}
describes both extreme cases. The sum over $K$ corresponds to $q=1$, while the minimum
results for $q \rightarrow -\infty$. Hence, a variation of the parameter $q$ allows to investigate
different possible scenarios lying between the best case and the worst case.  In order to model
synergy effects between different forces, one would have to add nonlinear terms, e.g. bilinear ones.
However, this would introduce a large number of additional parameters, which are even harder
to estimate than the first-order effects included in our model. Note that we already have non-linearities
in our model, namely the products $|P_i(\tau)| N_k(\tau)A_k{}^l(\tau)$ and the function $G_q$.
\par
The framework of supply networks also allows one to move from the semi-quantitative description 
of disaster management in Secs.~\ref{Kausal} to \ref{Dyn} to a fully quantitative one, if the required 
data are available (while we  work with assumption (\ref{approx}) otherwise):
One simple case of the supply network model proposed in Refs.~\cite{Production,Supplynet,stabil,Seba}
corresponds to the dynamic input-output model
\begin{equation}
\frac{dN_i }{d\tau} = \sum_j (\delta_{ij} - c_{ij} ) Q_j(\tau)
\end{equation}
with
\begin{equation}
 \frac{dQ_j}{d\tau} = \frac{V_j( N_j ) - Q_j(\tau)}{T_j} \, ,
\end{equation}
where $N_i $ denotes the inventory (stock level) of factor or product $i$, $V_j(N_j)$ the
desired and $Q_j\ge 0$ the actual throughput of sector $j$, $T_j$ the adaptation time, and 
$c_{ij}\ge 0$ the quantity of factor $i$ needed per throughput cycle. (For details see Ref.~\cite{Supplynet}.)
One could also say, $c_{ij}$ are the entries of the input-output matrix measured in economics,
and $c_{ij}Q_j$ is the flow of the quantity generated by factor $i$ to factor $j$.
In the limit of short adaptation times $T_j \approx 0$, the above equations reduce to
\begin{equation}
\frac{dN_i }{d\tau} = \sum_j (\delta_{ij} - c_{ij} ) V_j(N_j) \, .
\label{init}
\end{equation}
Let us assume that the stationary state of this supply system is given by $N_i(t) = N_i^0$. Moreover
let us denote the deviations from the stationary state by $\delta N_i(t) = N_i(t) - N_i^0$. With
\begin{equation}
 V_j(N_j) \approx V_j(N_j^0) + \frac{dV_j(N_j^0)}{dN_j} \, \delta N_j = A_j - B_j \, \delta N_j \, ,
\end{equation} 
the linearized version of Eq.~(\ref{init}) reads
\begin{equation}
 \frac{d\,\delta N_i}{d\tau} = \sum_j (W_{ij} - B_j \delta_{ij})\, \delta N_j(\tau)  \, ,
\label{toge2}
\end{equation}
where $A_j = V_j(N_j^0)$, $B_j = -dV_j(N_j^0)/dN_j >0$, and $W_{ij} = c_{ij} B_j$. Here, we have used that,
for the stationary solution $N_j^0$,
\begin{equation}
\sum_j (\delta_{ij} - c_{ij} ) ( A_j - B_j N_j^0) = 0 \, .
\end{equation}
In Eq.~(\ref{toge2}), we have $B_i = \sum_j W_{ji}$ because of $\sum_j c_{ji} = 1$. Taking into account the
additional contribution (\ref{toge}), we finally obtain the set of linear equations
\begin{equation}
 \frac{d\vec{P}}{d\tau} = (\mathbf{W} - \mathbf{D}) \vec{P}(\tau) = \mathbf{L} \vec{P}(\tau)
\end{equation}
with $\mathbf{W} = (W_{ij})$, $\mathbf{D}=(\delta_{ij} (B_i+D_i))$, 
$\mathbf{L} = \mathbf{W} - \mathbf{D}$, and $\vec{P}(\tau) = (P_i(\tau)) = (\delta N_i(\tau))$. 
Although $P_j$ can become negative due to some catastrophic impact,
Eqs.~(\ref{zwei}) and (\ref{drei}) for the average occurence times and their variance still remain meaningful,
when $D_i$ is replaced by $(B_i+D_i)$. Hence, they can be used to estimate the time of 
impact on other factors or sectors in the supply network.
\par
One particularly important aspect of supply networks is their sensitivity or robustness with respect to
perturbations. It is, for example, known that supply chains may suffer from the
so-called bullwhip effect, i.e. small temporal variations in the demand may cause large variations in the
supply. This instability leads both to undesireable delays in delivery at some places and large stock levels
at others \cite{Production,Supplynet,Dag}. In disaster management, this effect can have serious consequences.
However, anticipation is known to efficiently stabilize the dynamics of supply networks \cite{stabil,Seba}.
As the formulas from Sec.~\ref{Dyn} can be used to estimate the approximate time
at which certain factors are likely to be affected, they can help to optimize the supply chain management,
in particular to stabilize the supply of forces and materials in time. It is certainly reasonable to
have forces available in time to fight the spreading of the disaster to other sectors, rather than sending them
all to the places which are already devastated. The philosophy is to reach an anticipative disaster
management rather than having a responsive one. To model anticipation, the term
$\sum_i P_i(\tau) V_{ik} N_k(\tau) A_k{}^l(\tau)$ has to be replaced by
\begin{equation}
\sum_i |P_i(\tau+\Delta \tau)| V_{ik} N_k(\tau) A_k{}^l(\tau) \, ,
\end{equation}
where $\Delta \tau$ denotes the anticipation time horizon.
\par
\begin{figure}
\begin{center}
\begin{picture}(6,12.5)(0,5)
\put(0.05,17){\includegraphics[width=5.8\unitlength,clip=]{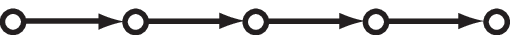}}
\put(0.05,15){\includegraphics[width=5.8\unitlength,clip=]{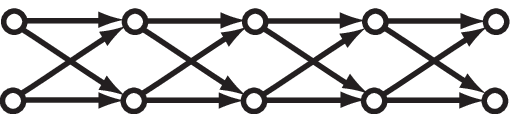}} 
\put(0.05,5.6){\includegraphics[width=5.8\unitlength,clip=]{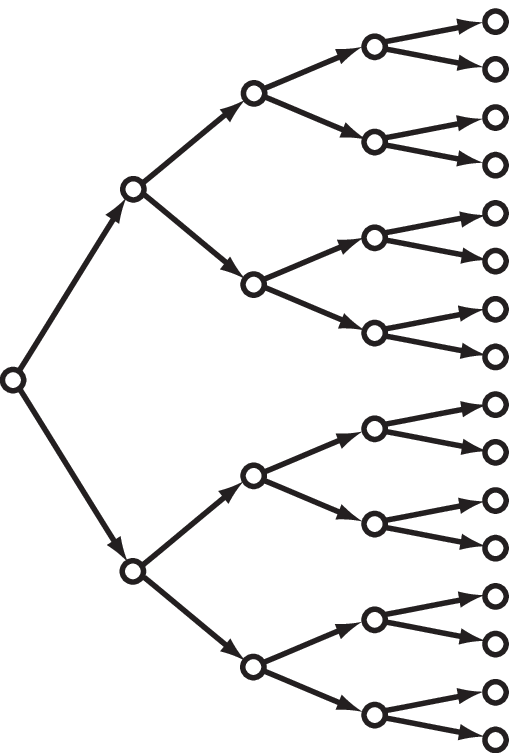}}
\put(0,17.5){{\footnotesize (a)}}
\put(0,16.5){{\footnotesize (b)}}
\put(0,12.1){{\footnotesize (c)}}
\end{picture}
\end{center}
\caption{Illustration of different supply networks: (a) linear supply chain, 
(b) ``supply ladder'', and (c) hierarchical supply tree.
The supply ladder is particularly robust because of its redundant links and nodes.} 
\label{SUP}
\end{figure}

Even more interesting is the robustness of supply networks with respect to structural changes,
e.g. when some supplier fails to work or to deliver. This may be investigated by changing the
coefficients $V_{ik}$ and $c_{iK}$, which characterize the supply network. The influence of
the topology of the supply network on its robustness, reliability, and dynamics is presently under
investigation \cite{Seba}. It has, for example, been noticed that supply ladders are more robust
than linear supply chains or supply hierarchies (see Fig.~\ref{SUP}). It is not surprising that
the redundance of supply ladders, i.e. the availability of alternative delivery channels, stabilizes
the system compared to a linear supply chain. We note, however, that hierarchical systems are
very common in disaster management, and better alternatives are expected to be found in
a research project that we presently pursue.
\par
\unitlength0.9cm
Some insights regarding the robustness of networks have already been gained for the world wide
web and other networks \cite{nets}. However, it is questionable whether the results for
small-world, scale-free or random networks can directly be transfered to disaster management.
Further research in this direction would be very helpful.

\section{Summary and Outlook} \label{summ}

In the past, physics has made significant contributions to the understanding of
catastrophes. This concerns, for example, the statistics of extreme events and
avalanche effects. In this paper, we have tried to indicate how physics could 
contribute to disaster management \cite{management}.
We have sketched a rather general, semi-quantitative approach for the assessment
of interaction networks which can serve for decision support, as it allows to compare 
potential optimization measures including their side effects. 
The method takes into account feedback loops and can be
generalized to a dynamical model, which is suitable to estimate likely sequences of
events and the times at which they are expected to materialize. We have also 
described the dynamics of disaster management in a way similar to excitable
media, distinguishing an excitable state (``ready to go''), an active state, and a refractory
(exhausted) state requiring recovery. As successful actions need the simultaneous presence
and/or action of several specialized forces and particular equipment, there was also
a direct connection with the management of supply networks. We have pointed out that
the management of disasters and supply chains can be considerably improved
by anticipation, based on the formulas in this paper. Moreover, the 
robustness  \cite{robustness} crucially depends on the structure of the supply network.
In an on-going study, we investigate the optimal network structure to achieve robustness 
with respect to dynamical and structural perturbations \cite{Seba}. The statistical physics
of networks and graphs \cite{nets} is expected to make significant contributions to this.
Some relevant aspects for the optimization of organizations and work groups
have already been studied \cite{organization}.
\par
Our proposed approach connects to several methods and fields from statistical physics, such as the
master equation \cite{stoch,Path}, excitable media \cite{excite},
and the dynamics of transport (supply) processes \cite{Supplynet}. It is also in the tradition of
system dynamics \cite{SystDyn}, which has, with some success, been used to anticipate 
future problems of society \cite{global}. In such kinds of studies, it is reasonable to 
carry out a sensitivity analysis \cite{sensitive} and to investigate the impact of random
effects \cite{stoch}, which do, of course, play a significant role for the dynamics of
catastrophes. Apart from stochastic methods, one may in the future also apply elements
of fuzzy logic \cite{fuzzy} in order to describe the vague knowledge and soft facts,
on which disaster management is often based. Insufficient, inconsistent, and uncertain
information is one of the typical complications of disaster management, which makes
it difficult to assess alternatives and to take the best decision, in particular under often
very tense time constraints. In the future, information theory \cite{information} is expected
to make some valuable contributions to the design of decision support systems which
can integrate inconsistent information and handle incomplete information \cite{NJP}. 
\par
In summary, statistical physics offers various promising concepts to develop and improve methods 
of disaster management. We think that the theory of self-organization \cite{selforg} is
particularly promising for this, having in mind principles such as synchronization
\cite{synchro},  distributed control \cite{distrib}, and optimal self-organization \cite{optim}.
It is expected that an application of these principles
would lead to a more flexible, efficient, and robust disaster management compared to
the present centralized or hierarchical concepts.

\section*{Acknowledgments}

This work has been inspired by interviews of several central
persons involved into the management of the disastrous floodings in
Germany during August 2002. We want to express our particular
thanks to the Major of Dresden, Ingolf Ro{\ss}berg, to the director of the South-East Branch of the
German Railway Net (DB Netz), Ralf Rothe, to the director of the
Dam Administration of the State of Saxony, Hans-J\"urgen Glasebach,
to the director of the Traffic Alliance Oberelbe (VVO),
Knut Ringat, to the managing director of the Dresdner Transport Services (DVB), Frank M\"uller-Eberstein,
to the Chief of the Fire Fighter Brigade of Pirna, Mr. Peter Kammel, and many others.

\end{document}